\documentclass{aastex62}

\usepackage{textcomp}
\usepackage[normalem]{ulem}
\newcommand{\sm}{$\sim$}

\usepackage{amsmath}

\usepackage[utf8]{inputenc}

\usepackage{hyperref}
\usepackage{multirow}
\usepackage{amsmath}
\usepackage{graphicx} 
\usepackage{xcolor}
\usepackage{url}

\usepackage[para,online,flushleft]{threeparttable}
\usepackage{hyperref}

\shorttitle{Identifying and Tracking Solar Magnetic Flux Elements with Deep Learning}
\shortauthors{Jiang et al.}

\begin{document}

\title{{\bf \large Identifying and Tracking Solar Magnetic Flux Elements with Deep Learning}}

\author{Haodi Jiang}
\affiliation{Institute for Space Weather Sciences, New Jersey Institute of Technology, 
University Heights, Newark, NJ 07102-1982, USA
	hj78@njit.edu, jw438@njit.edu,  chang.liu@njit.edu, ju.jing@njit.edu, hl422@njit.edu, wangj@njit.edu, haimin.wang@njit.edu}
\affiliation{Department of Computer Science, New Jersey Institute of Technology, University Heights, Newark, NJ 07102-1982, USA}

\author{Jiasheng Wang}
\affiliation{Institute for Space Weather Sciences, New Jersey Institute of Technology, University Heights, Newark, NJ 07102-1982, USA
	hj78@njit.edu, jw438@njit.edu,  chang.liu@njit.edu, ju.jing@njit.edu, hl422@njit.edu, wangj@njit.edu, haimin.wang@njit.edu}
\affiliation{Center for Solar-Terrestrial Research, New Jersey Institute of Technology, University Heights, Newark, NJ 07102-1982, USA}
\affiliation{Big Bear Solar Observatory, New Jersey Institute of Technology, 40386 North Shore Lane, Big Bear City, CA 92314-9672, USA}

\author{Chang Liu}
\affiliation{Institute for Space Weather Sciences, New Jersey Institute of Technology, University Heights, Newark, NJ 07102-1982, USA
		hj78@njit.edu, jw438@njit.edu,  chang.liu@njit.edu, ju.jing@njit.edu, hl422@njit.edu, wangj@njit.edu, haimin.wang@njit.edu}
\affiliation{Center for Solar-Terrestrial Research, New Jersey Institute of Technology, University Heights, Newark, NJ 07102-1982, USA}
\affiliation{Big Bear Solar Observatory, New Jersey Institute of Technology, 40386 North Shore Lane, Big Bear City, CA 92314-9672, USA}

\author{Ju Jing}
\affiliation{Institute for Space Weather Sciences, New Jersey Institute of Technology, University Heights, Newark, NJ 07102-1982, USA
	hj78@njit.edu, jw438@njit.edu,  chang.liu@njit.edu, ju.jing@njit.edu, hl422@njit.edu, wangj@njit.edu, haimin.wang@njit.edu}
\affiliation{Center for Solar-Terrestrial Research, New Jersey Institute of Technology, University Heights, Newark, NJ 07102-1982, USA}
\affiliation{Big Bear Solar Observatory, New Jersey Institute of Technology, 40386 North Shore Lane, Big Bear City, CA 92314-9672, USA}

\author{Hao Liu}
\affiliation{Institute for Space Weather Sciences, New Jersey Institute of Technology, University Heights, Newark, NJ 07102-1982, USA
	hj78@njit.edu, jw438@njit.edu,  chang.liu@njit.edu, ju.jing@njit.edu, hl422@njit.edu, wangj@njit.edu, haimin.wang@njit.edu}
\affiliation{Department of Computer Science, New Jersey Institute of Technology, University Heights, Newark, NJ 07102-1982, USA}

\author{Jason T. L. Wang}
\affiliation{Institute for Space Weather Sciences, New Jersey Institute of Technology, University Heights, Newark, NJ 07102-1982, USA
	hj78@njit.edu, jw438@njit.edu,  chang.liu@njit.edu, ju.jing@njit.edu, hl422@njit.edu, wangj@njit.edu, haimin.wang@njit.edu}
\affiliation{Department of Computer Science, New Jersey Institute of Technology, University Heights, Newark, NJ 07102-1982, USA}

\author{Haimin Wang}
\affiliation{Institute for Space Weather Sciences, New Jersey Institute of Technology, University Heights, Newark, NJ 07102-1982, USA
	hj78@njit.edu, jw438@njit.edu,  chang.liu@njit.edu, ju.jing@njit.edu, hl422@njit.edu, wangj@njit.edu, haimin.wang@njit.edu}
\affiliation{Center for Solar-Terrestrial Research, New Jersey Institute of Technology, University Heights, Newark, NJ 07102-1982, USA}
\affiliation{Big Bear Solar Observatory, New Jersey Institute of Technology, 40386 North Shore Lane, Big Bear City, CA 92314-9672, USA}

\begin{abstract}
Deep learning has drawn a lot of interest in recent years due to its effectiveness 
in processing big and complex observational data gathered from diverse instruments. 
Here we propose a new 
deep learning method, called SolarUnet, 
to identify and track solar magnetic flux elements or features in observed vector magnetograms
based on the Southwest Automatic Magnetic Identification Suite (SWAMIS). 
Our method consists of a data pre-processing component that prepares training data
from the SWAMIS tool,
a deep learning model implemented as a U-shaped convolutional neural network for fast and accurate image segmentation, 
and a post-processing component that prepares tracking results.
SolarUnet is applied to data from the 1.6 meter Goode Solar Telescope 
at the Big Bear Solar Observatory.  
When compared to the widely used SWAMIS tool,
SolarUnet is faster while agreeing mostly with SWAMIS on feature size and flux distributions, 
and complementing SWAMIS in tracking  long-lifetime features.
Thus, the proposed physics-guided deep learning-based tool can be considered as an 
alternative method for 
solar magnetic tracking.
\end{abstract}


\section{Introduction} 
\label{sec:intro}
Tracking magnetic flux elements
is an important subject in heliophysics research 
\citep{DeForest_2007,Leenaarts_2015,Wang_2018}.\footnote{In the study presented here, we focus on tracking
signed, including positive and negative, magnetic flux elements.}
Identifying and tracking the surface
magnetic elements is useful in deriving statistical parameters of the local and global solar dynamo, 
allowing for sophisticated analyses of solar activity \citep{DeForest_2007}. 
It not only helps scientists understand the distribution of magnetic fluxes \citep{intro1}, 
but also helps estimate the amount of energy in acoustic waves, 
which play an important part in the heating of the solar chromosphere and corona \citep{Fossum_2006}. 
In addition, magnetic tracking is useful in deriving boundary conditions of 
magnetohydrodynamic (MHD) modeling of the solar corona and solar wind. 
In the past, many researchers have studied the behaviors and patterns of magnetic flux elements.
For instance, \citet{Chen_2015} developed a technique to detect and classify small-scale magnetic flux cancellations 
and link them to chromospheric rapid blueshifted excursions. 
\citet{2018A&A...611A..56G} investigated the occurrence and persistence of magnetic elements in the quiet Sun 
to understand the scales of organization at which turbulent convection operates. 
\citet{2018ApJ...859L..26M} reported findings related to small-scale magnetic flux emergence in the quiet Sun. 

In magnetic tracking, features are defined as a visually identifiable part of an image, 
such as a clump of magnetic flux or a blob in a magnetogram. 
One of the most popular software tools for magnetic feature tracking 
across multiple images/frames
is the Southwest Automatic Magnetic Identification Suite \citep[SWAMIS;][]{DeForest_2007}. 
SWAMIS takes five steps to track magnetic flux elements: 
(1) feature discrimination for each frame; 
(2) feature identification within a frame; 
(3) feature association across frames; 
(4) occasional noise filtering; 
and (5) event detection \citep{Chen_2015}. 
Magnetic events are broadly classified into two categories: death and birth \citep{Lamb_2008}; 
the former refers to the end of a magnetic feature’s existence while
the latter refers to the start of a magnetic feature’s existence. 

In this paper, we present a new tool, called SolarUnet, to track magnetic flux elements.
Our tool is built using deep learning \citep{LeCun2015}.
The tool can detect three different types of events in each category, 
namely (i) disappearance and appearance,  
(ii) merging and splitting, 
and (iii) cancellation and emergence. 
The event ``disappearance" is defined as the end of a single unipolar magnetic feature 
that ``fades away" to nothing in the absence of nearby features 
across two frames; 
the opposite event ``appearance" is defined as the origin of a single unipolar feature 
where the unipolar feature does not exist in the previous frame.
The event ``merging" is defined as 
the combination of two or more like-sign features into a single magnetic feature;
the opposite event  ``splitting" is defined as 
the breakup of a single magnetic feature into at least two like-sign features, 
where the total flux of all child features is roughly the same as that of the parent feature.
The event ``cancellation" is defined as the demise of a magnetic feature that collides with one or more opposite-sign features,
resulting in the demise of these features or an alive feature carrying the remaining flux;
the event ``emergence" is defined as the appearance of  
opposite-sign features with approximately the same magnitude 
or a new feature adjacent to previously existing opposite-sign features
in a nearly flux-conserving manner.

Deep learning, which is a subfield of machine learning, has drawn a lot of interest in recent years \citep{LeCun2015}.
Inspired by its success in computer vision, speech recognition and natural language processing, 
researchers have started to use deep learning in astronomy and astrophysics 
\citep{Huertas_Company_2018, 10.1093/mnras/sty3217, Kim2019, 10.1093/mnras/stz761, 2019ApJ...877..121L, 10.1093/mnras/stz333}.
In contrast to the existing methods for magnetic tracking 
\citep{Lamb_2010, Lamb_2013, Chen_2015}, 
our SolarUnet tool is built using deep learning.
Compared to the most closely related magnetic tracking tool, SWAMIS, 
which uses hysteresis as the discrimination scheme and a 
gradient-based ``downhill" method to identify features in a frame,
SolarUnet runs faster while producing similar or complementary results.

The rest of this paper is organized as follows.
Section 2 describes observations and data used in this study.
Section 3 presents details of SolarUnet and tracking algorithms used by the tool.
Section 4 reports experimental results.
Section 5 concludes the paper.

\section{Observations and Data Preparation}
\label{sec:observational data}

We adopted two collections of observations in this study.
The first collection was conducted by
the Near InfraRed Imaging Spectropolarimeter \citep[NIRIS;][]{2012ASPC..463..291C}
of the 1.6 m Goode Solar Telescope (GST) at the Big Bear Solar Observatory  
\citep[BBSO;][]{2010AN....331..636C, 2010ApJ...714L..31G, 2012SPIE.8444E..03G, 2014SPIE.9147E..5DV}.
This collection contained 
observations of the magnetic polarity inversion region in NOAA AR 12665 (431\arcsec, $-$131\arcsec) 
during \sm20:16--22:42 UT on 2017 July 13. 
The obtained data included spectro-polarimetric observations of a full set of Stokes measurements at the Fe I 1564.8 nm line 
(0.25 $\r{A}$
bandpass) by NIRIS with a FOV (field of view) of 80\arcsec\ at 0\arcsec.24 resolution and 56 s cadence.
Vector magnetic field products in local coordinates were constructed 
after removing azimuth ambiguity \citep{2009ASPC..415..365L}. 

The second collection of observations was conducted with a clear seeing condition; 
BBSO/GST achieved diffraction-limited imaging during \sm16:17--22:17 UT
on 2018 June 07. 
The obtained multi-wavelength observations revealed detailed structural and evolutionary properties 
of small scale magnetic polarities in quiescent solar regions at north of the disk center  ($-$32\arcsec, 294\arcsec). 
The essential data included in this collection were the images taken by the GST's NIRIS using a 2048$\times$2048 pixels Teledyne camera 
with a \sm80\arcsec\ FOV. 
The spatial resolution (at diffraction limit $\theta= \lambda / D$) of the NIRIS images was 0$\farcs$2, 
and the temporal cadence was 56~s. 
The magnetograms were then aligned based on sunspot and plage features, 
with an alignment accuracy within 0\arcsec.3, which was the best accuracy by using interpolation.  

We prepared our training and testing sets by using the magnetograms 
taken from the two collections of observations described above. 
Because the magnetograms taken on 2018 June 07 had higher quality than the observations conducted on 2017 July 13, 
we used the higher-quality magnetograms to prepare our training data so as to obtain a better magnetic tracking model. 
Specifically, we gathered all 202 frames from the second collection of observations, 
and excluded 6 images with poor quality (these excluded images were very noisy). 
The remaining 196 frames were used as training data for magnetic tracking.
The testing set contained all 147 magnetograms from the first collection of observations. 
Table \ref{table_samples} summarizes the numbers of training and testing images used in this study.

\begin{table}[t]
\centering
\caption{Numbers of Images Used in Our Study}
 \setlength\tabcolsep{20.0pt}
\label{table_samples}
\begin{tabular}{c c}
\hline\hline    
Number of Training Images  & Number of Testing Images   \\ [1ex]  
\hline
196 (from the second collection)  & 147  (from the first collection)  \\
\hline
\end{tabular}
\end{table}

\newpage
\section{Methodology} 
\label{sec:method}

\subsection{Overview of SolarUnet}
\label{sec:overview}

\begin{figure}
	\epsscale{1.05}
	\plotone{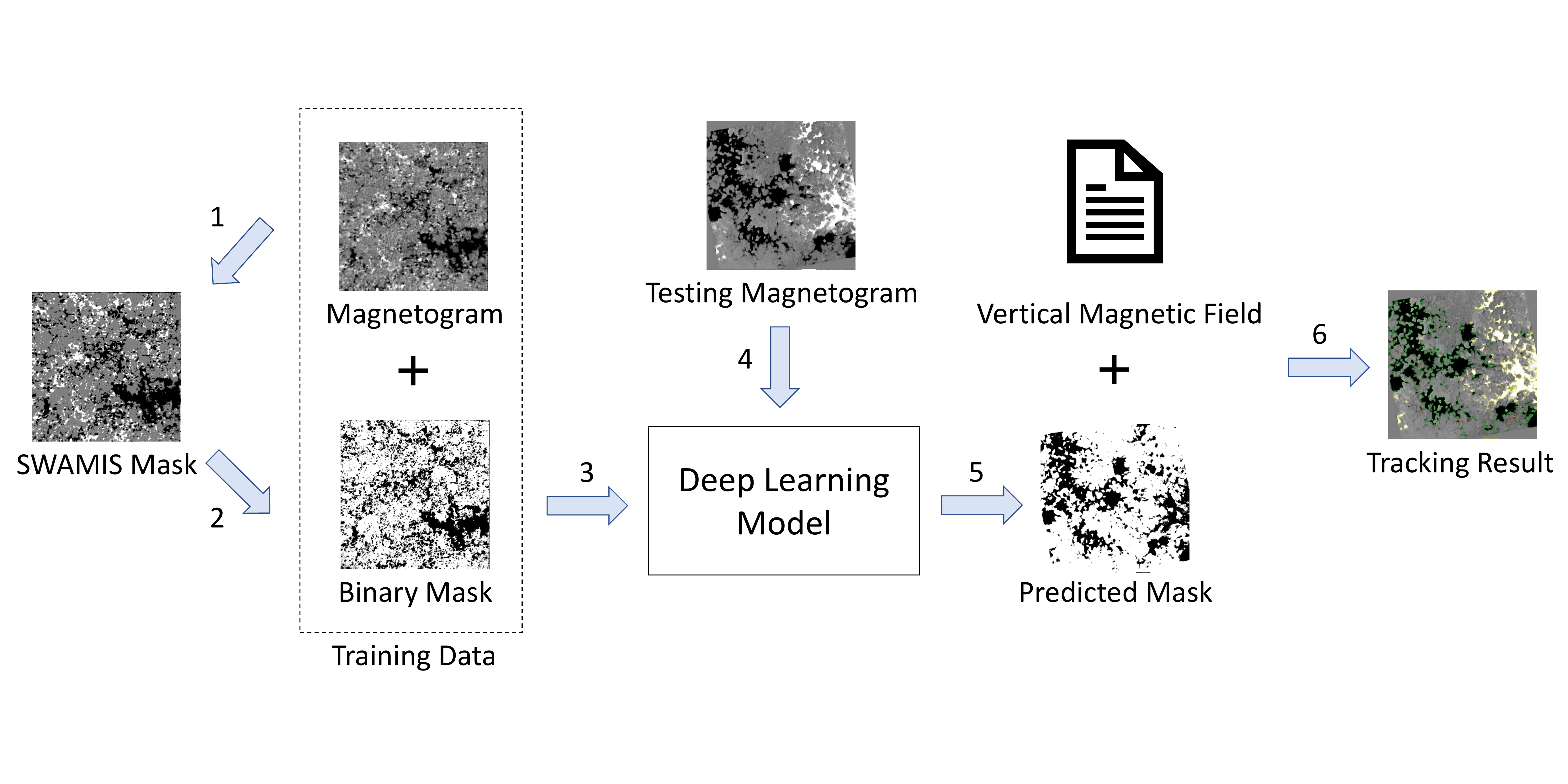}
	\caption{Illustration of the proposed method (SolarUnet) for identifying and tracking solar magnetic flux elements.
		SolarUnet employs a deep learning model for image segmentation.
		The training data used to train the deep learning model are highlighted in the dashed box.} 
	\label{fig:method}
\end{figure}

Figure \ref{fig:method} explains how SolarUnet works.
Training magnetograms are pre-processed in steps 1 and 2, 
and then used to train the deep learning model for image segmentation (step 3).
The trained model takes a testing magnetogram (step 4) and produces a predicted mask (step 5).
Through post-processing of the predicted mask, SolarUnet produces magnetic tracking results (step 6).

Specifically, in step 1, we apply SWAMIS with the downhill option
to the 196 training magnetograms to get 196 masks.
These images, including the magnetograms and masks, 
are converted to 8-bit grayscale images of 720$\times$720  pixels, 
suitable for our deep learning model.
Pixels in the masks belong to three classes represented by three colors/labels respectively:  
positive magnetic flux with a label of 1 (white), 
negative magnetic flux with a label of $-$1 (black), and 
non-significant flux with a label of 0 (gray). 
During pre-processing, we convert the 196 three-class masks obtained from SWAMIS
to 196 two-class (binary) masks by 
(i) changing the label of the non-significant flux regions from 0 to 1;
and (ii) changing both the positive magnetic flux regions and negative magnetic flux regions to
significant flux regions with label $-$1 (step 2).

The 196 magnetograms (images) and two-class (binary) masks 
are then used to train the deep learning model, implemented in 
TensorFlow \citep{tensorflow2015-whitepaper} and Keras \citep{chollet2015keras},
for image segmentation (step 3).
Because our deep learning model needs a large amount of data in order to train successfully,
the model invokes the ImageDataGenerator\footnote{\url{https://www.tensorflow.org/api_docs/python/tf/keras/preprocessing/image/ImageDataGenerator}} 
in Keras to perform data augmentation, expanding the 
training set by shifting, rotating, flipping and scaling the training images during the model training process.
Shifting an image is to move all pixels of the image
horizontally or vertically while keeping the dimensions of the image the same. 
Rotating an image is to rotate the image clockwise by a given number of degrees from 0 to 360.
Flipping an image is to reverse the rows or columns of pixels in the image.
Scaling an image is to randomly zoom the image in and either add new pixel values around the image or interpolate pixel values in the image.
We train the deep learning model using 1 epoch with 10,000 iterations/epoch.
In each iteration, the model randomly selects one of the 196 training magnetograms and its binary mask,
feeds them to the ImageDataGenerator to generate a synthetic magnetogram and binary mask, 
and uses the synthetic magnetogram and binary mask to train the model.
There are 10,000 iterations and hence 10,000 synthetic magnetograms and binary masks are generated through the data augmentation process,
where the 10,000 generated magnetograms and binary masks 
are used for model training.\footnote{Notice that SWAMIS is applied only to the 196 training magnetograms
mentioned in Table \ref{table_samples}; SWAMIS is never run on the 10,000 generated (synthetic) magnetograms.}
We have chosen to use data augmentation as opposed to acquiring more training data because 
the quality of ground-based observations is subject to many factors such as seeing conditions and observing time limits. 
Obtaining large volumes of high-quality training data requires further observations. 
Nevertheless, using the synthetic training images produces reasonably good results as shown in Section \ref{sec:experiment}.

When a testing magnetogram is submitted, it is converted to a 8-bit grayscale image of 720$\times$720 pixels, 
and fed to the trained deep learning model (step 4).
The trained deep learning model predicts a two-class (binary) mask, 
containing  non-significant flux regions with label 1 and
significant flux regions with label $-$1 (step 5). 
We convert the predicted two-class (binary) mask back to a three-class mask 
via post-processing as follows.
For the non-significant flux regions with label 1, we change their label from 1 to 0.
For the significant flux regions with label $-$1, we use the information of radial components 
in the vertical magnetic field image shown in Figure \ref{fig:method},
where the radial components are perpendicular to the plane of the Sun, 
to reconstruct positive and negative magnetic flux regions.  
Specifically, for each pixel $x$ in the significant flux regions in the predicted binary mask, 
we check the magnetic strength of the pixel, $y$, at $x$'s corresponding location in the vertical magnetic field image.
If $y$'s magnetic strength is greater than 150 G,
we set $x$ as positive magnetic flux and change the label of this pixel from $-$1 to 1. 
If $y$'s  magnetic strength is smaller than $-$150 G, 
we set $x$ as negative magnetic flux and the label of this pixel remains $-$1.
If $y$'s  magnetic strength is between $-$150 G and 150 G,
we set $x$ as non-significant flux and change the label of this pixel from $-$1 to 0.
This yields a three-class mask with the polarity information.

Finally, we apply our magnetic tracking algorithms described in Section \ref{sec:magnetic tracking} 
to the testing magnetogram and masks
to get tracking results  (step 6).
Magnetic tracking is often involved with more than one testing magnetogram, 
and we output the tracking results in all of the testing magnetograms.

\subsection{Implementation of the Deep Learning Model in SolarUnet}
\label{sec:implementation}

\begin{figure}
	\epsscale{1.15}
	\plotone{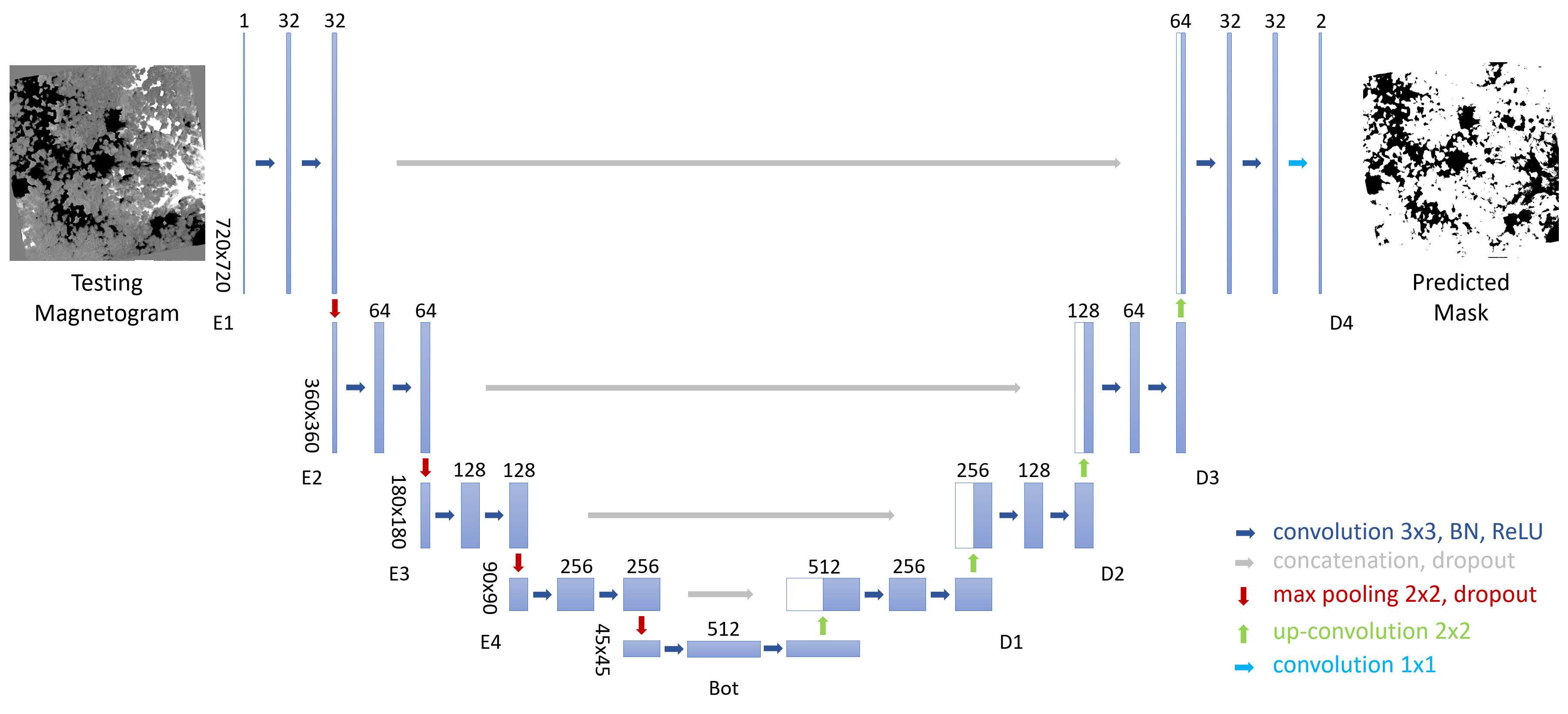}
	\caption{Illustration of the deep learning model used in SolarUnet.
          This model is a U-shaped convolutional neural network, consisting of an encoder, a bottleneck, a decoder, followed by a pixelwise binary classification layer.
          The encoder is comprised of 4 blocks: E1, E2, E3, E4. 
          Each block has two 3$\times$3 convolution layers, represented by blue arrows, 
	followed by a 2$\times$2 max pooling layer, represented by a red arrow. 
	 The decoder is also comprised of 4 blocks: D1, D2, D3, D4. 
	Each block has two 3$\times$3 convolution layers followed by a 2$\times$2 up-convolution layer, represented by a green arrow. 
	The bottleneck, denoted Bot, mediates between the encoder and the decoder. 
	It uses two 3$\times$3 convolution layers followed by a 2$\times$2 up-convolution layer.}
	\label{fig:my_unet}
\end{figure}

Figure \ref{fig:my_unet} illustrates the deep learning model used in SolarUnet, 
which is a U-shaped convolutional neural network.
We adapt U-Net \citep{U-Net-2019} to our work, enhancing it to obtain our model.
The model has an encoder, a bottleneck, a decoder, followed by a pixelwise binary classification layer.\footnote{Please see the Appendix for
more detailed descriptions of the technical terms used here.}
The encoder consists of 4 blocks: E1, E2, E3, E4. 
Each block has two 3$\times$3 convolution layers, represented by blue arrows, 
followed by a 2$\times$2 max pooling layer, represented by a red arrow. 
In each convolution layer, we adopt batch normalization \citep[BN;][]{bn} after convolution, 
followed by a rectified linear unit (ReLU) activation function.
Furthermore, we add a dropout layer \citep{dropout} after each max pooling layer.
The four encoder blocks E1, E2, E3, E4 have 32, 64, 128 and 256 kernels, respectively.

The bottleneck, denoted Bot, mediates between the encoder and the decoder. 
It uses two 3$\times$3 convolution layers followed by a 2$\times$2 up-convolution layer, represented by a green arrow. 
The bottleneck has 512 kernels.  
Similar to the encoder, the decoder consists of 4 blocks: D1, D2, D3, D4. 
Each block has two 3$\times$3 convolution layers followed by a 2$\times$2 up-convolution layer. 
The four decoder blocks D1, D2, D3, D4 have 256, 128, 64 and 32 kernels, respectively. 
The input of each decoder block is concatenated by the output of the corresponding encoder block
where the concatenation is represented by a gray arrow. 
A dropout layer is added after each concatenation.
Finally, a 1$\times$1 convolution layer, represented by a turquoise arrow, with 2 kernels 
followed by a softmax activation function,
is used to produce a segmentation mask. 
During testing, the deep learning model
takes as input a testing magnetogram and produces as output a two-class mask.

The input resolution of the encoder block E1 is set to 720$\times$720 pixels to match the size of the testing magnetogram. 
Each max pooling layer reduces the size by a factor of 2.
Hence, the input resolution of the encoder block E2 (E3, E4 respectively) is
360$\times$360 (180$\times$180, 90$\times$90 respectively) pixels.
The input resolution of the bottleneck, Bot, is 45$\times$45 pixels.
Each up-convolution layer increases the size by a factor of 2.
Thus, the input resolution of the decoder block D1 (D2, D3, D4 respectively) is 
90$\times$90 (180$\times$180, 360$\times$360, 720$\times$720 respectively) pixels.

The loss function, $L$, used by the deep learning model is 
the binary cross-entropy function defined below:
\begin{equation}
L = - \sum_{x} \log y_{c}(x, W).
\label{magnetic loss function}
\end{equation}
Here, $W$ are the parameters of the convolutional neural network,
$y_{c}(x, W)$ is the output of the softmax layer of the convolutional neural network, 
and $c$ is the class label (1 vs. $-$1) of each pixel $x$.

In training the deep learning model, we adopt  adaptive moment estimation (Adam) 
to find the optimal parameters of the model.
The learning rate of Adam is set to 0.0001. 
Adam combines the advantages of two popular methods: AdaGrad and RMSProp \citep{Goodfellow-et-al-2016}. 
In most cases, Adam achieves better performance than other stochastic optimization methods 
including the stochastic gradient descent (SGD) with momentum employed by U-Net \citep{Goodfellow-et-al-2016}.

Although both our deep learning model and U-Net \citep{U-Net-2019} have the same U-shaped architecture, 
they differ in several ways.
First, U-Net used SGD with a momentum of 0.99 to train and optimize its model. 
By contrast, we choose Adam because it achieves better performance in our case where 
the training process would be trapped in a poor local minimum if SGD were used. 
Second, U-Net focused on imbalanced datasets and used a weighted cross-entropy loss function 
to tackle the imbalanced classification problem. 
By contrast, because our training set is relatively balanced in the sense that 
non-significant flux regions roughly have the same number of pixels as significant flux regions, 
we use the binary cross-entropy loss function as defined in Eq. (\ref{magnetic loss function}). 
Third, we adopt batch normalization and dropout layers, which were not used by U-Net. 
Batch normalization improves model learning, 
stabilizes the learning process, reduces the learning (training) time and improves prediction accuracy \citep{bn}. 
Dropout prevents neural networks from overfitting \citep{dropout}, where
overfitting means that a trained model fits training data too well, and cannot generalize 
to make predictions on unseen testing data.
Finally, we reduce the numbers of kernels of the encoder, bottleneck and decoder blocks 
by a factor of 2 compared to U-Net to speed up the training process and reduce GPU memory usage.

\clearpage
\subsection{Algorithms for Magnetic Tracking and Event Detection}
\label{sec:magnetic tracking}

After describing the deep learning model used in SolarUnet,
we now turn to the magnetic tracking algorithms employed by SolarUnet.
Based on the positive magnetic flux regions and negative magnetic flux regions found in Section \ref{sec:overview},
we identify signed magnetic flux elements or features 
in a magnetogram (image/frame)
by utilizing a connected-component labeling algorithm \citep{He:2009:FCL:1542560.1542851} 
to group all adjacent segments in the
positive magnetic flux regions and negative magnetic flux regions respectively
 if their pixels in edges or corners touch each other.
We filter out those magnetic features whose sizes are smaller than a user-determined threshold.
The features eliminated from consideration are treated as noise.
Then, we assign each of the remaining features a label number 
and highlight the features with different bordering colors. 
Finally, we consider the association of features (magnetic flux elements) across different frames 
to perform event detection.

Based on the observational data and instruments used, 
we calculate the moving distance $D$ (number of pixels) of a magnetic flux element $X$ as follows:
\begin{equation}
D = \dfrac{C \times \mbox{cadence}}{725 \hspace*{+0.1cm} km/arcsec \times \Delta s}
\label{moving-distance}
\end{equation}
where $C$ is the transverse speed ($km$/s) on the photosphere 
according to the observational environment and Sun's activity, and
$\Delta$s is the pixel scale.
In this study, $C$ is set to 4 $km$/s.
For the NIRIS magnetograms used here, $\Delta$s = 0.083\arcsec /pixel.
We then calculate the radius of the region of interest (ROI) with respect to the location or position of the magnetic flux element $X$, 
denoted ROI$_{p(X)}$, as follows:
	\begin{equation}
radius(\mbox{ROI}_{p(X)}) = 2 \times D + r
\label{ROI}
	\end{equation}
where $r$ is the radius of the smallest region that covers the magnetic flux element.
The ROI$_{p(X)}$ defines the region which the magnetic flux element $X$ can not move beyond between two contiguous frames.

The magnetic flux $\varPhi (X)$ is calculated by the surface integral of 
the normal component of magnetic field passing through $X$, as follows: 
\begin{equation}
\varPhi (X) = \int_{S} B_{z} \,dS
\label{flux-formula}
\end{equation}
where $B_{z}$ is the magnitude of the magnetic field from the vertical magnetic field image of the testing magnetogram and
$S$ is the area of the surface of $X.$
	
For any two features or magnetic flux elements $X$ and $Y$ in a frame, 
we define the distance between $X$ and $Y$, denoted $Dist(X, Y)$, as follows:
\begin{equation}
Dist(X, Y) = \min_{x \in X, y \in Y} d(x, y)
\end{equation}
where $x$ and $y$ are pixels in $X$ and $Y$ respectively, 
and $d(x, y)$ is the Euclidean distance between $x$ and $y$.
For a given magnetic flux element $X$ in a frame, 
the adjacent features of $X$ are defined as
the $k$-nearest neighboring features of $X$ in the frame.
(In the study presented here, $k$ is set to 10.)

Let $X_{i}$ be a magnetic flux element in the current frame F$_{1}$.
Let $Y_{i}$ be a magnetic flux element in the next frame F$_{2}$
where $Y_{i}$ occurs in the ROI$_{p(X_{i})}$ in F$_{2}$.
We say $X_{i}$ is approximately equal to $Y_{i}$, denoted $X_{i} \approx Y_{i}$, if 
$X_{i}$ and $Y_{i}$ have the same sign, and
\begin{equation}
|\frac{\varPhi(X_{i})-\varPhi(Y_{i})}{\varPhi(X_{i})}| \leq \epsilon_{1}
\end{equation}
where $\epsilon_{1}$ is a user-determined threshold based on the observation setting and tracking task requirement. 
(In the study presented here, $\epsilon_{1}$ is set to 0.33.)

With the above terms and definitions, we are now ready to describe our algorithms for
magnetic tracking and event detection. 
For each magnetic flux element or feature $X_{i}$ in the current frame F$_{1}$,
the algorithms below determine and indicate whether $X_{i}$ 
exists in the next frame F$_{2}$, or $X_{i}$ is involved in a merging or cancellation event, or
$X_{i}$ disappears. \\
\  \\
$[$Main Algorithm$]$ \\
\indent \indent (\romannumeral 1) If there exists 
a magnetic feature $Y_{i}$ in the ROI$_{p(X_{i})}$ in the next frame F$_{2}$
such that $X_{i}$ $\approx$ $Y_{i}$, then indicate $X_{i}$ exists in F$_{2}$
(more precisely, $X_{i}$ becomes $Y_{i}$ in F$_{2}$)
and go to (\romannumeral 2); otherwise go to (\romannumeral 3).\\
\indent \indent (\romannumeral 2) Check the sign of $X_{i}$, highlighting $X_{i}$ by yellow bordering color if $X_{i}$ is positive
and by green bordering color if $X_{i}$ is negative. Exit the Main Algorithm.\\
\indent \indent (\romannumeral 3) Find all magnetic features
in the ROI$_{p(X_{i})}$ in the current frame F$_{1}$.
Group those features in the ROI$_{p(X_{i})}$ whose signs are the same as the sign of $X_{i}$
into $G_{same}(X_{i})$
and group those features in the ROI$_{p(X_{i})}$ whose signs are opposite to the sign of $X_{i}$
into $G_{opposite}(X_{i})$. \\
\indent \indent (\romannumeral 4) Go to the Merging Algorithm to check whether 
$X_{i}$ and the features in $G_{same}(X_{i})$ meet the merging criterion.
If yes, perform the merging using the Merging Algorithm and then exit the Main Algorithm; 
otherwise indicate $X_{i}$ is not involved in a merging event.\\
\indent \indent (\romannumeral 5) Go to the Cancellation Algorithm to check whether
$X_{i}$ and the features in $G_{opposite}(X_{i})$ meet the cancellation criterion.
If yes, perform the cancellation using the Cancellation Algorithm and then exit the Main Algorithm; 
otherwise indicate $X_{i}$ is not involved in a cancellation event.\\
\indent \indent (\romannumeral 6) If $X_{i}$ is not involved in a merging event according to (\romannumeral 4) and 
$X_{i}$ is not involved in a cancellation event based on (\romannumeral 5), indicate $X_{i}$ disappears and highlight 
$X_{i}$ by purple bordering color.\footnote{ For the events belonging to the death category, 
namely disappearance, merging and cancellation, magnetic features involved in the events are highlighted by different bordering colors
(purple for disappearance, amber for merging and pink for cancellation)
 in the current frame  F$_{1}$.
For the events belonging to the birth category, 
namely appearance, splitting and emergence, magnetic features involved in the events are highlighted by different bordering colors
(blue for appearance, aqua for splitting and red for emergence)
 in the next frame  F$_{2}$.}
Exit the Main Algorithm.\\
\  \\
$[$Merging Algorithm$]$ \\
\indent \indent (\romannumeral 1)  For each feature $X_{j}$
in $G_{same}(X_{i})$,  check whether there exists a feature $Y_{j}$
in the ROI$_{p(X_{j})}$ in the next frame F$_{2}$ such that $X_{j}$ $\approx$ $Y_{j}$, and
if yes, delete $X_{j}$ from $G_{same}(X_{i})$.
Call the remaining set, $G'_{same}(X_{i})$. 
If there are too many features in $G'_{same}(X_{i})$, only keep those adjacent features of $X_{i}$ in $G'_{same}(X_{i})$.\\
\indent \indent (\romannumeral 2) If there exist 
a combination $C_{s}$ of features in $G'_{same}(X_{i})$ and
a magnetic feature $Y_{i}$ in the ROI$_{p(X_{i})}$
in the next frame F$_{2}$ where $Y_{i}$ and $X_{i}$ have the same sign, 
such that Eq. (\ref{diffm}) below is satisfied, then we
say $X_{i}$ and the features in $C_{s}$ are merged into $Y_{i}$:
\begin{equation}
|\frac{(\varPhi(X_{i}) + \sum_{X \in C_{s}} \varPhi(X)) - \varPhi(Y_{i})}{\varPhi(Y_{i})}| \leq \epsilon_{2}
\label{diffm}
\end{equation}
where $\varPhi(X_{i})$ and $\varPhi(X)$ have the same sign,
$\epsilon_{2}$ is a user-determined threshold (which is set to 0.5).
Indicate $X_{i}$ and the features in $C_{s}$ are merged into $Y_{i}$ by
highlighting $X_{i}$ and the features in $C_{s}$ using amber bordering color.
Exit the Merging Algorithm. \\
\indent \indent (\romannumeral 3)
If there does not exist $Y_{i}$ or a combination of features satisfying Eq. (\ref{diffm})
(i.e., the condition in (ii) is not satisfied),
indicate $X_{i}$ and the features in $G_{same}(X_{i})$ do not meet the merging criterion.
Exit the Merging Algorithm. \\
\  \\
$[$Cancellation Algorithm$]$ \\
\indent \indent (\romannumeral 1)  For each feature $X_{j}$
in $G_{opposite}(X_{i})$,  check whether there exists a feature $Y_{j}$
in the ROI$_{p(X_{j})}$ in the next frame F$_{2}$ such that $X_{j}$ $\approx$ $Y_{j}$, and
if yes, delete $X_{j}$ from $G_{opposite}(X_{i})$.
Call the remaining set, $G'_{opposite}(X_{i})$. 
If there are too many features in $G'_{opposite}(X_{i})$, only keep those adjacent features of $X_{i}$ in $G'_{opposite}(X_{i})$.\\
\indent \indent (\romannumeral 2)
If there exists a combination $C_{o}$ of features in $G'_{opposite}(X_{i})$ 
such that Eq. (\ref{diffc1}) below is satisfied, then we say $X_{i}$ and the features in $C_{o}$ cancel each other
(referred to as balanced cancellation in \citet{DeForest_2007}):
\begin{equation}
|\frac{\varPhi(X_{i}) + \sum_{X \in C_{o}} \varPhi(X) }{\varPhi(X_{i})}| \leq \epsilon_{2}
\label{diffc1}
\end{equation}
where $\varPhi(X_{i})$ and $\varPhi(X)$ have opposite signs.
Indicate $X_{i}$ and the features in $C_{o}$ cancel each other by
highlighting $X_{i}$ and the features in $C_{o}$ using pink bordering color. 
Exit the Cancellation Algorithm. \\
\indent \indent (\romannumeral 3) If there exist a combination $C_{o}$ of features in $G'_{opposite}(X_{i})$ 
and a magnetic feature $Y_{i}$ in the ROI$_{p(X_{i})}$ in the next frame F$_{2}$, 
such that Eq. (\ref{diffc2}) below is satisfied, 
then we say $X_{i}$ and the features in $C_{o}$ are canceled to yield a feature $Y_{i}$ carrying the remaining flux
(referred to as unbalanced cancellation in \citet{DeForest_2007}):
\begin{equation}
	\frac{\left| |\varPhi(X_{i}) + \sum_{X \in C_{o}} \varPhi(X)| - |\varPhi(Y_{i})| \right|}{|\varPhi(Y_{i})|} \leq \epsilon_{2}
	\label{diffc2}
\end{equation}
where $\varPhi(X_{i})$ and $\varPhi(X)$ have opposite signs.
Indicate $X_{i}$ and the features in $C_{o}$ are canceled by
highlighting $X_{i}$ and the features in $C_{o}$ using pink bordering color.
Exit the Cancellation Algorithm. \\
\indent \indent (\romannumeral 4)
If neither the condition in (ii) nor the condition in (iii) is satisfied, 
indicate $X_{i}$ and the features in $G_{opposite}(X_{i})$ do not meet the cancellation criterion.
Exit the Cancellation Algorithm. \\
\  \\
To determine whether the magnetic feature $X_{i}$ appears 
or is involved in a splitting or emergence event, we use the same algorithms as described above
except that we treat the next frame F$_{2}$ as the current frame and
the current frame F$_{1}$ as the next frame.

\section{Results} \label{sec:experiment}

SolarUnet is implemented in Python.\footnote{\url{https://www.python.org/}}
Our deep learning model is coded with TensorFlow\footnote{\url{https://www.tensorflow.org/}} \citep{tensorflow2015-whitepaper} 
and Keras\footnote{\url{https://keras.io/}} \citep{chollet2015keras} libraries.
The data processed by SolarUnet, with the aid of Astropy\footnote{\url{https://www.astropy.org/}} \citep{2013A&A...558A..33A},
 include FITS files containing vector magnetic fields, 
PNG images of the observational data described in Section \ref{sec:observational data}, 
and image masks obtained from the SWAMIS tool presented in \citet{DeForest_2007}.
SWAMIS, written in PDL (Perl Data Language)\footnote{\url{http://pdl.perl.org/}} and
available via SolarSoft\footnote{\url{https://sohowww.nascom.nasa.gov/solarsoft/}}  
\citep{solarsoft-1998}, was run with the downhill option.
Figures in this section were produced with the aid of matplotlib\footnote{\url{https://matplotlib.org/}} \citep{Hunter:2007}. 
Statistical tests were performed by SciPy\footnote{\url{https://www.scipy.org/}} \citep{2020SciPy-NMeth}.
All experiments were conducted on a Dell PC with i7-8700k CPU, 32GB RAM and a NVIDIA GeForce RTX 2080 GPU 
for training and testing the deep learning model. 

\subsection{Magnetic Tracking and Event Detection Results}

In this series of experiments, we used the 196 magnetograms mentioned in Table \ref{table_samples} 
and the corresponding masks obtained from SWAMIS to train SolarUnet as described in Section \ref{sec:overview},
and then we performed testing on the set of 147 magnetograms mentioned in Table \ref{table_samples}.
The testing set contained observations in NOAA AR 12665 (431\arcsec, $-$131\arcsec) during \sm 20:16--22:42 UT on 2017 July 13. 
The filter size threshold was fixed at 10 pixels.
Thus, in the experiments we considered features or magnetic flux elements having at least 10 pixels.

We present figures to illustrate the six events studied here.
Frames taken at 20:15:49 UT and 20:16:45 UT are used to illustrate a disappearance event.
Frames taken at 21:00:48 UT and 21:01:45 UT are used to illustrate an appearance event. 
Frames taken at 20:18:38 UT and 20:19:34 UT are used to illustrate a merging event.
Frames taken at 20:19:34 UT and 20:20:30 UT are used to illustrate a splitting event and a cancellation event.
Frames taken at 20:17:41 UT and 20:18:38 UT are used to illustrate an emergence event. 
We present enlarged FOV results in these figures with a FOV of 7.5\arcsec\
where each figure has two axes: E-W (x-axis) and S-N (y-axis).

\begin{figure}
	\epsscale{0.8}
	\plotone{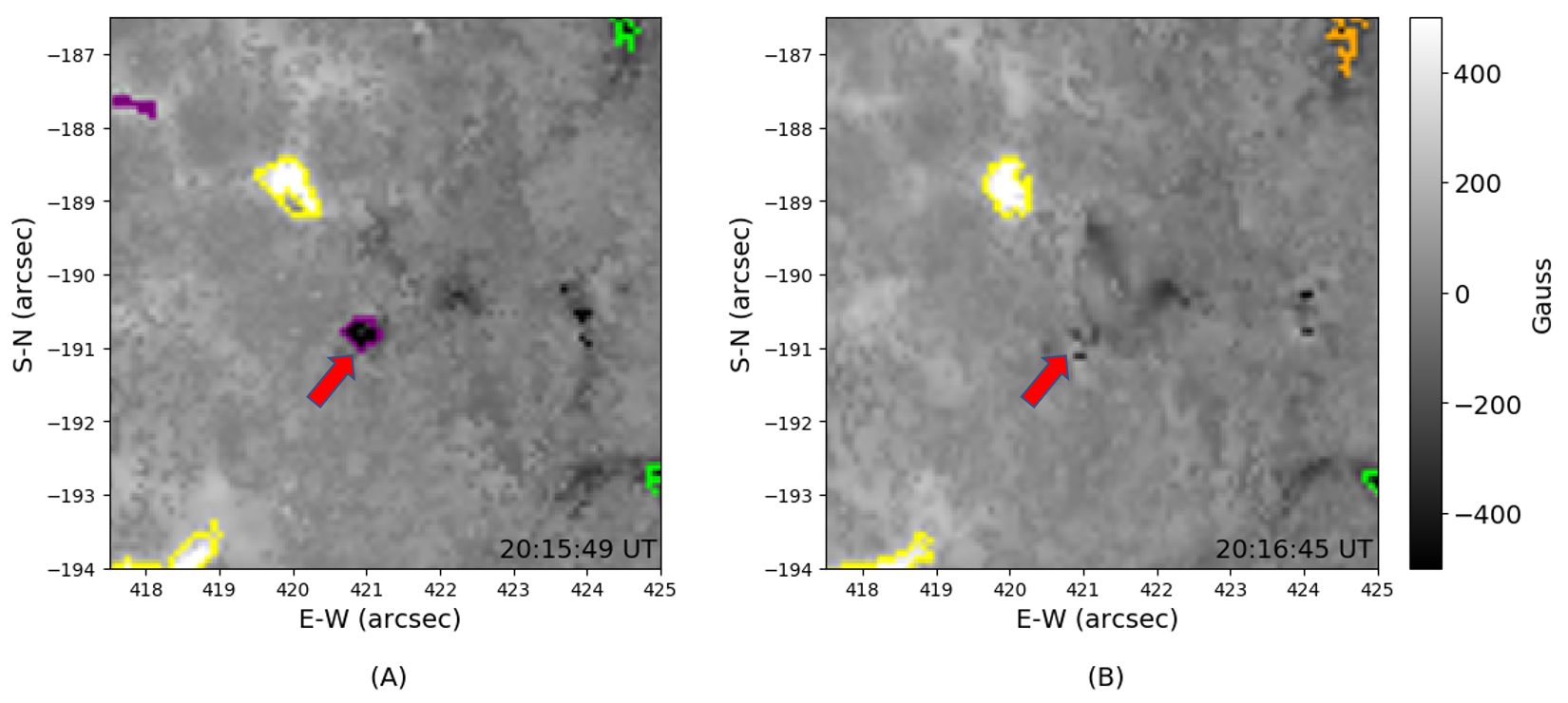}
	\caption{Example of BBSO/GST images of a disappearance event. 
The negative magnetic flux element highlighted by purple bordering color in (A) disappears in (B).
Time in UT is at the bottom right of each image.}
	\label{fig:disappearance}
\end{figure}

\begin{figure}
	\epsscale{0.8}
	\plotone{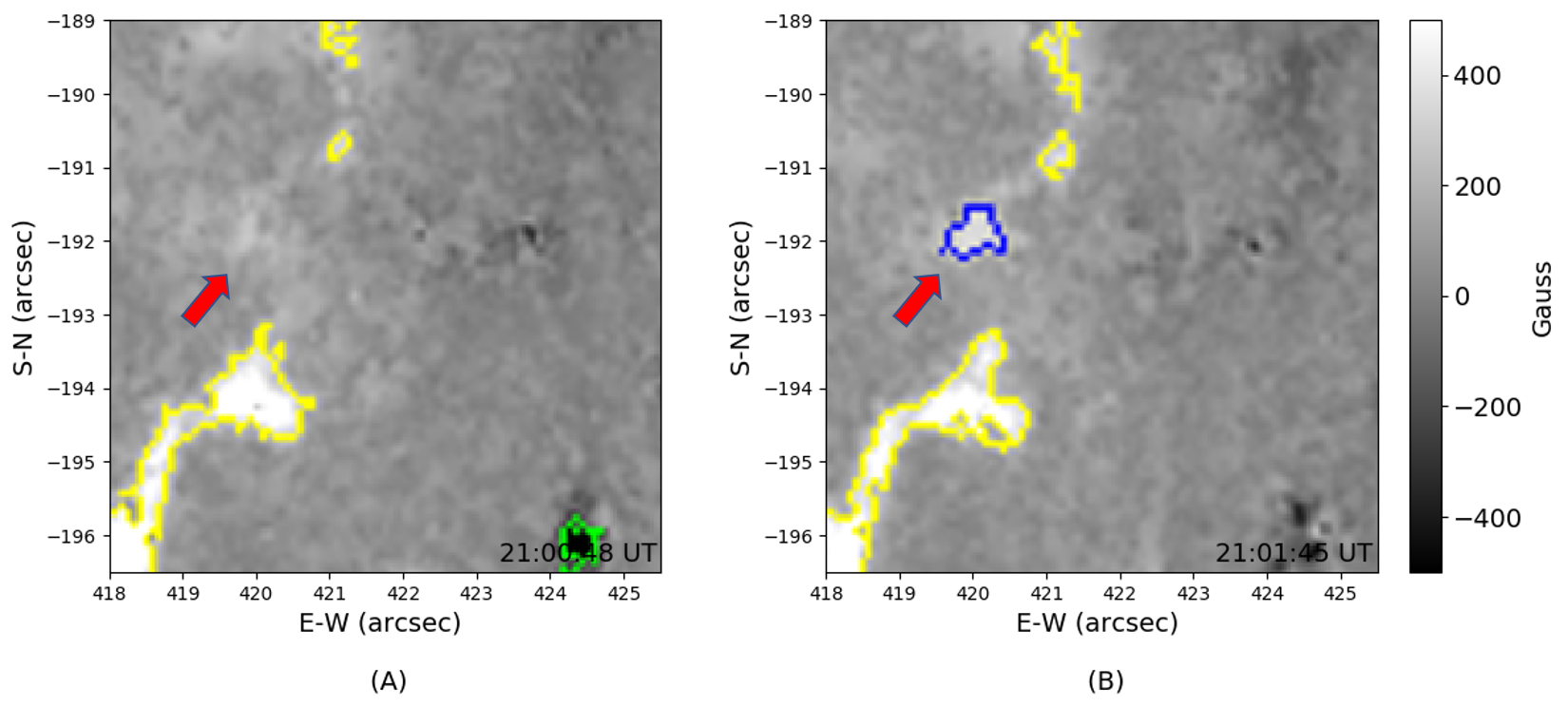}
	\caption{Example of BBSO/GST images of an appearance event.
The positive magnetic flux element highlighted in blue bordering color in (B) does not exist in (A),
and hence an appearance event is detected.
Time in UT is at the bottom right of each image.}
	\label{fig:appearance}
\end{figure}

Figure \ref{fig:disappearance} shows a disappearance event.
In Figure \ref{fig:disappearance}(A), a magnetic feature highlighted by purple bordering color exists at
 E-W = 421\arcsec\ and S-N = $-191$\arcsec,
which is pointed to by a red arrow
in the frame from 20:15:49 UT.
This feature disappears in the next frame from 20:16:45 UT as shown in Figure \ref{fig:disappearance}(B).
Figure \ref{fig:appearance} illustrates an appearance event.
In Figure \ref{fig:appearance}(A), there exists no feature at
E-W = 420\arcsec\ and S-N = $-192$\arcsec\ in the frame from 21:00:48 UT.
However, a new feature appears in the next frame from 21:01:45 UT,
which is highlighted by blue bordering color and pointed to by a red arrow as shown in Figure \ref{fig:appearance}(B).

\begin{figure}
	\epsscale{0.8}
	\plotone{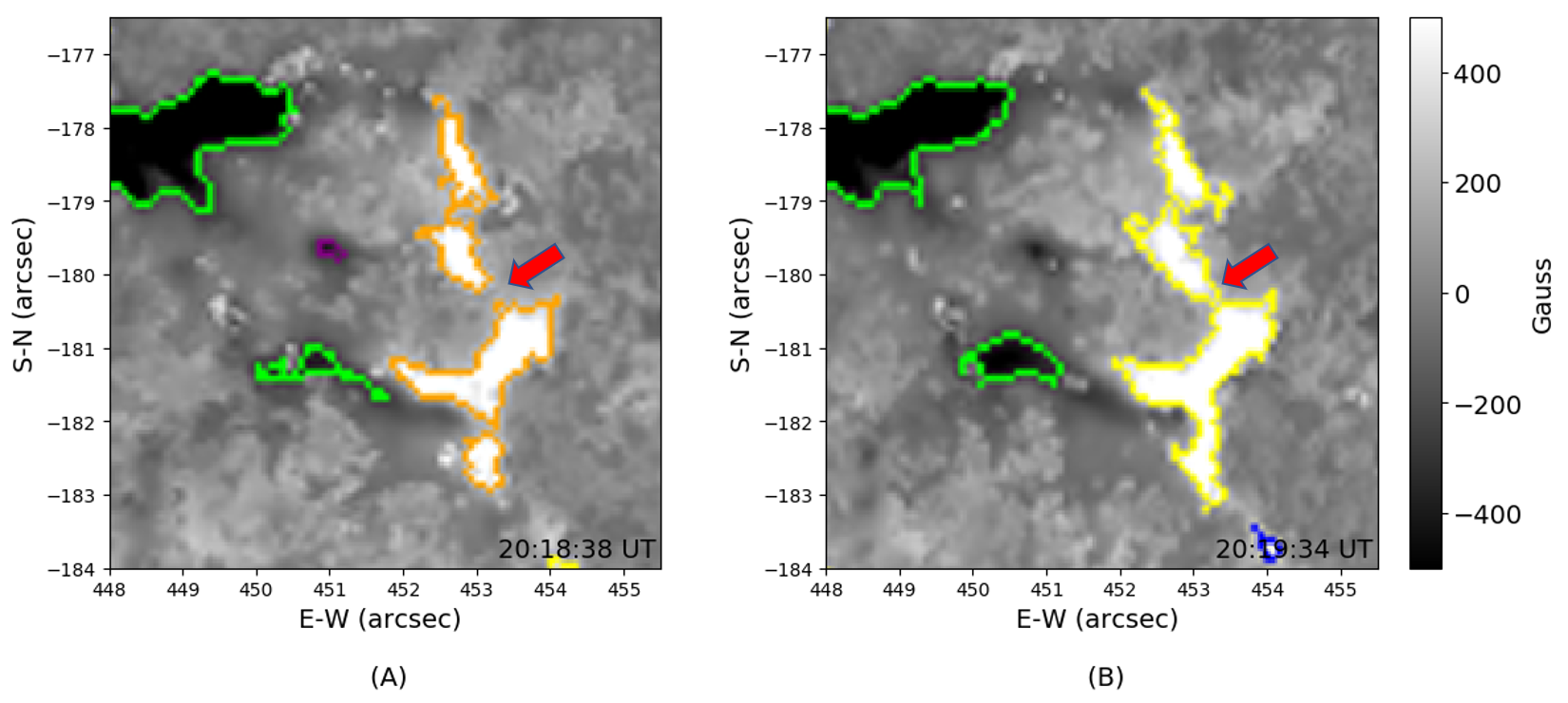}
	\caption{Example of BBSO/GST images of a merging event.
Two positive magnetic flux elements highlighted by amber bordering color in (A) are merged into a single
positive magnetic flux element in (B). 
Time in UT is at the bottom right of each image.}
	\label{fig:merging}
\end{figure}

\begin{figure}
	\epsscale{0.8}
	\plotone{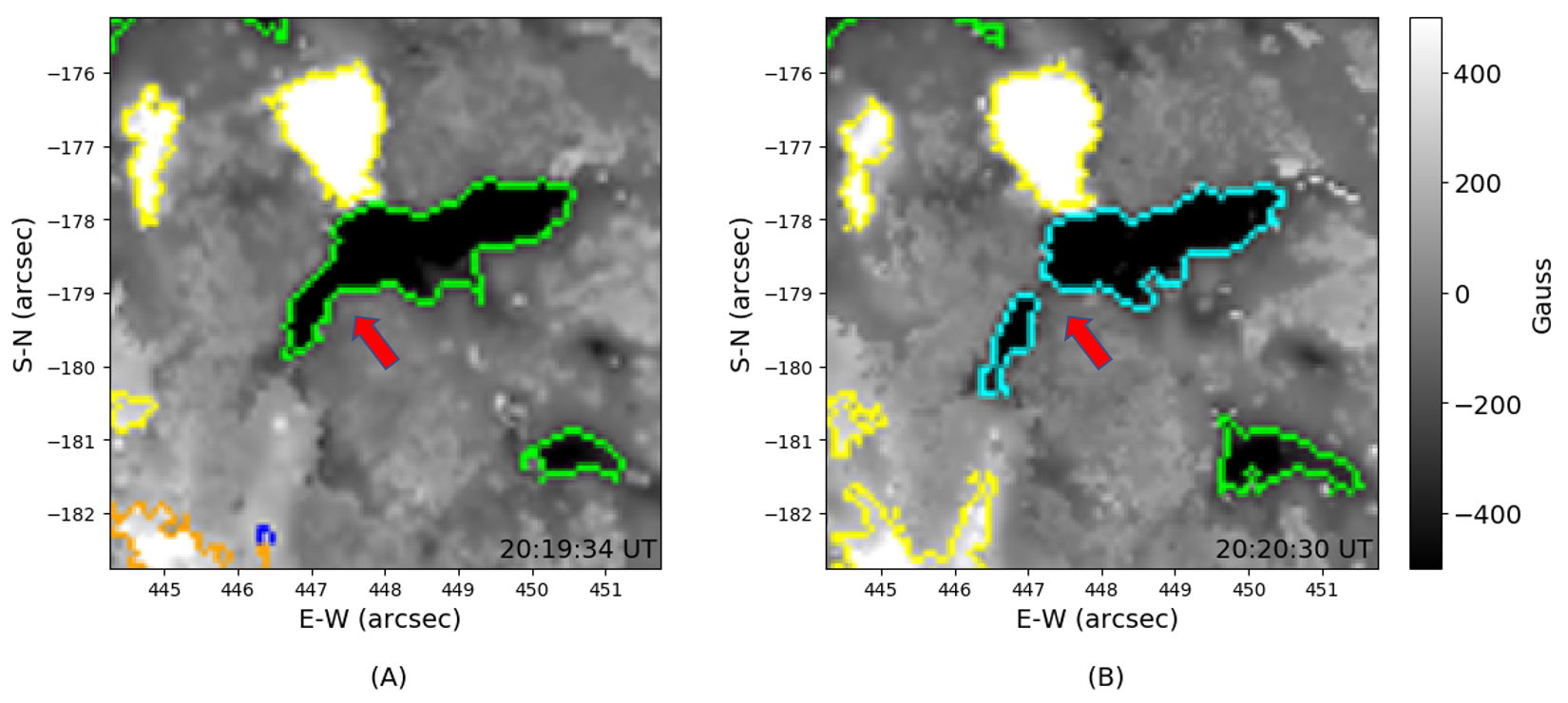}
	\caption{Example of BBSO/GST images of a splitting event.
A negative magnetic flux element in (A) is split into two negative magnetic flux elements highlighted by
aqua bordering color in (B).
Time in UT is at the bottom right of each image.}
	\label{fig:splitting}
\end{figure}

Figure  \ref{fig:merging} shows a merging event.
In Figure \ref{fig:merging}(A), 
two separate positive polarity magnetic features highlighted by amber bordering color and pointed to by a red arrow 
in the frame from 20:18:38 UT
are merged into a single positive polarity feature at
E-W = 453\arcsec\ and S-N = $-180$\arcsec\ in the frame from 20:19:34 UT as shown in Figure  \ref{fig:merging}(B).
Figure \ref{fig:splitting} illustrates a splitting event.
In Figure \ref{fig:splitting}(A), a negative polarity feature exists at
E-W = 448\arcsec\ and S-N = $-179$\arcsec\ in the frame from 20:19:34 UT.
This negative polarity feature is split into two negative polarity features highlighted by aqua bordering color and 
pointed to by a red arrow in the frame from 20:20:30 UT as shown in Figure \ref{fig:splitting}(B).

\begin{figure}
	\epsscale{0.8}
	\plotone{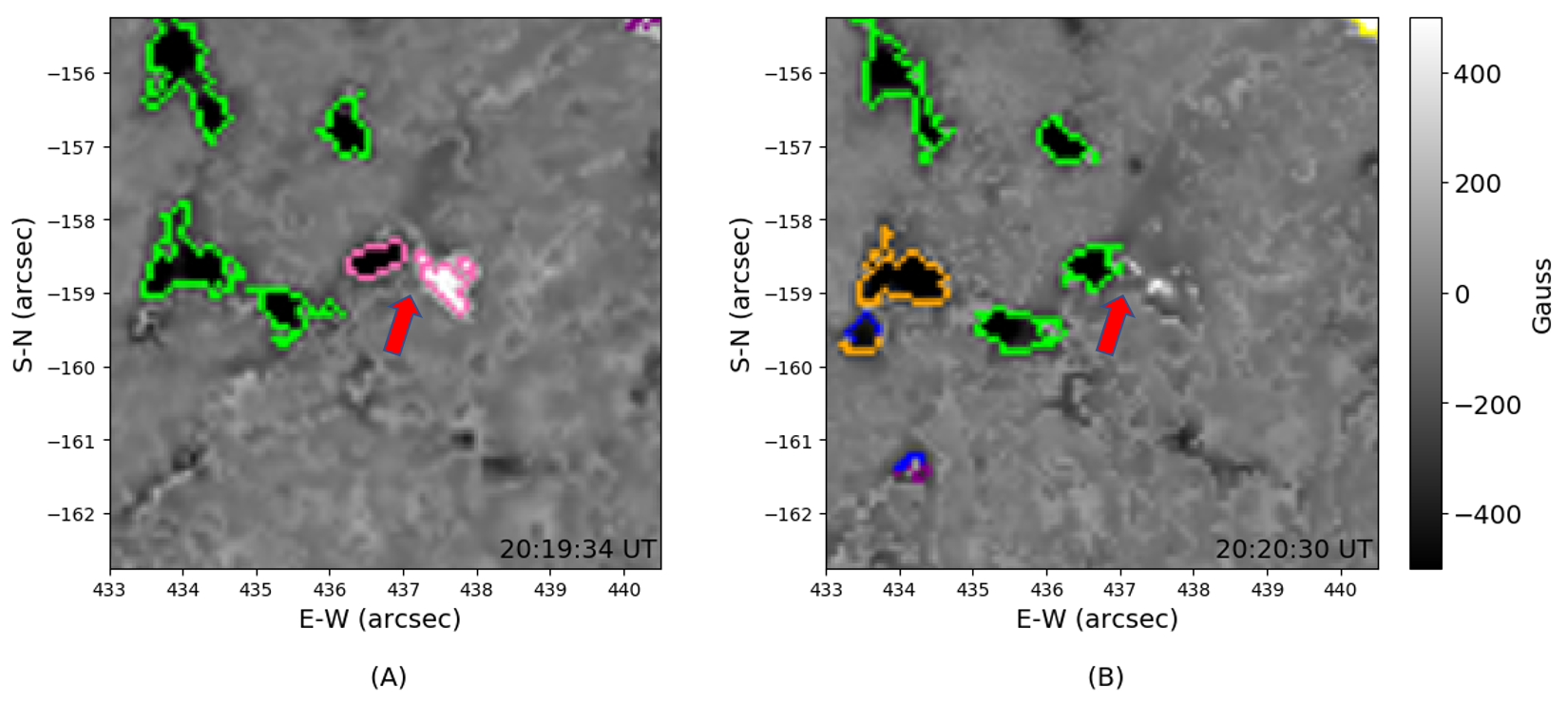}
	\caption{Example of BBSO/GST images of an unbalanced cancellation event. 
A positive magnetic flux element and a negative magnetic flux element,
both of which are highlighted by pink bordering color in (A), are canceled to yield
a negative magnetic flux element carrying the remaining flux, which is pointed to by a red arrow in (B).
Time in UT is at the bottom right of each image.}
	\label{fig:cancellation}
\end{figure}

\begin{figure}
	\epsscale{0.8}
	\plotone{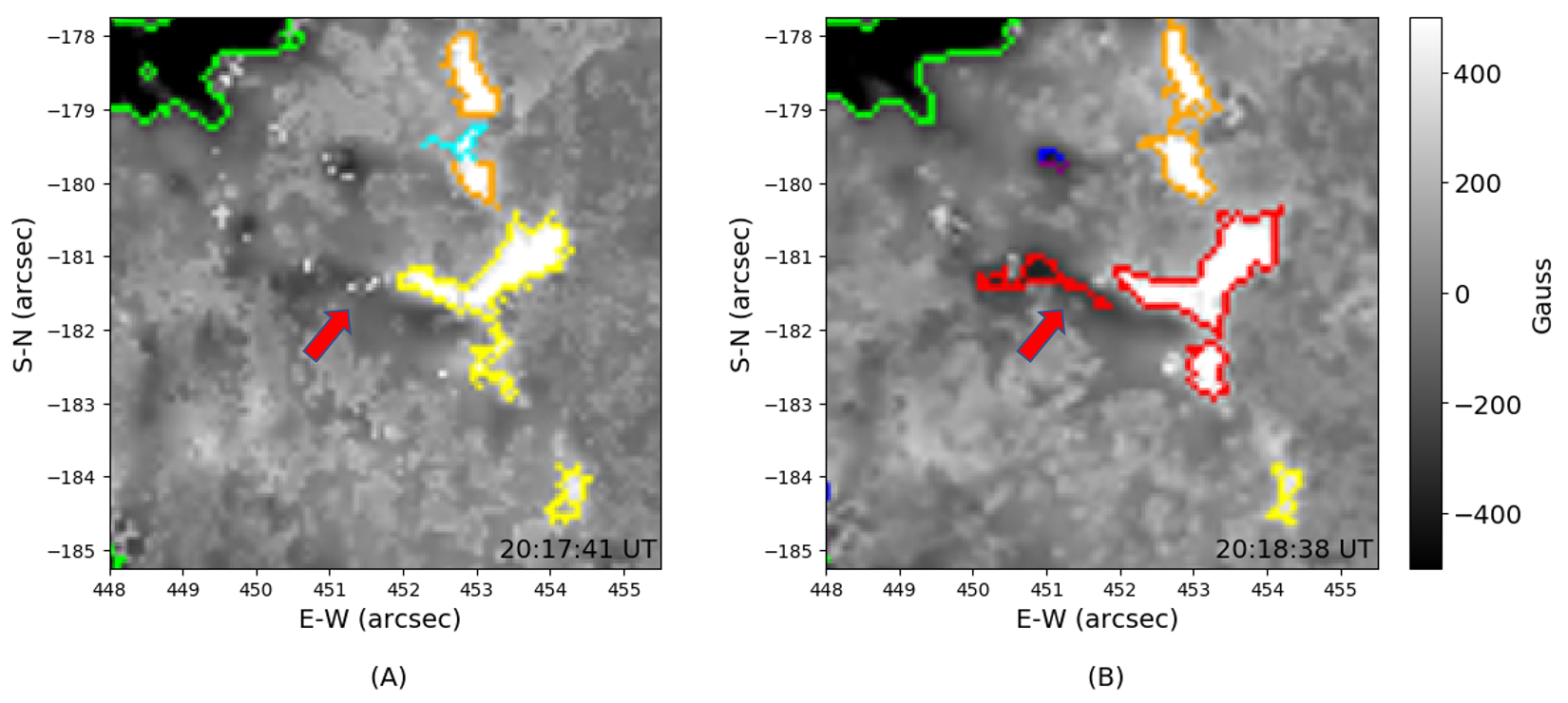}
	\caption{Example of BBSO/GST images of an unbalanced emergence event. 
A new negative magnetic flux element emerges, next to a pre-existing positive magnetic flux element in (A), 
in a nearly flux-conserving manner
where the two magnetic flux elements with opposite signs are highlighted by red bordering color in (B).
Time in UT is at the bottom right of each image.}
	\label{fig:emergence}
\end{figure}

Figure \ref{fig:cancellation} shows an unbalanced cancellation event.
In Figure \ref{fig:cancellation}(A), there exist two magnetic features with opposite signs around
E-W = 437\arcsec\ and S-N = $-158.5$\arcsec\ in the frame from 20:19:34 UT.
The two magnetic features with opposite signs, highlighted by pink bordering color,
are canceled to yield a negative polarity magnetic feature carrying the remaining flux, pointed to by a red arrow, 
in the frame from 20:20:30 UT as shown in Figure \ref{fig:cancellation}(B).
Figure \ref{fig:emergence} illustrates an unbalanced emergence event.
A new negative polarity feature emerges, next to a pre-existing positive polarity feature, 
in the frame from 20:18:38 UT as shown in Figure \ref{fig:emergence}(B).
The flux of the positive polarity feature pointed to by a red arrow in Figure \ref{fig:emergence}(A) is approximately
equal to the total flux of the two features with opposite signs, highlighted 
by red bordering color and  pointed to by a red arrow, in Figure \ref{fig:emergence}(B).

\clearpage
\subsection{Comparison with SWAMIS}

While both SolarUnet and SWAMIS \citep{DeForest_2007} 
aim to track magnetic features and detect magnetic events,
they differ in two ways.
\begin{enumerate}
\item
Their feature discrimination and identification algorithms are different.
SWAMIS used hysteresis and a threshold-based method to separate non-significant flux regions, 
positive magnetic flux regions and negative magnetic flux regions.
Then it used direct clumping and a gradient based (``downhill'') method to identify magnetic features in these regions.
By contrast, SolarUnet employs a U-shaped convolutional neural network to
gain knowledge from training data, and then predicts a binary (two-class) mask
containing non-significant flux regions and significant flux regions.
Next, SolarUnet separates the significant flux regions
into positive magnetic flux regions and negative magnetic flux regions
through post-processing of the binary mask.
Finally, SolarUnet uses a connected-component labeling algorithm \citep{He:2009:FCL:1542560.1542851} 
to group all adjacent segments in the
positive magnetic flux regions and negative magnetic flux regions respectively
if their pixels in edges or corners touch each other to 
identify positive and negative magnetic flux elements.
\item
Their feature tracking and event detection algorithms are different.
SWAMIS used a dual-maximum-overlap criterion to find persistent features across frames. 
In contrast, SolarUnet defines the region of interest (ROI) of a magnetic feature and
traces the flux changes of the magnetic features in the ROI to find the association
of features across frames.
\end{enumerate}

\begin{figure}
	\epsscale{0.8}
	\plotone{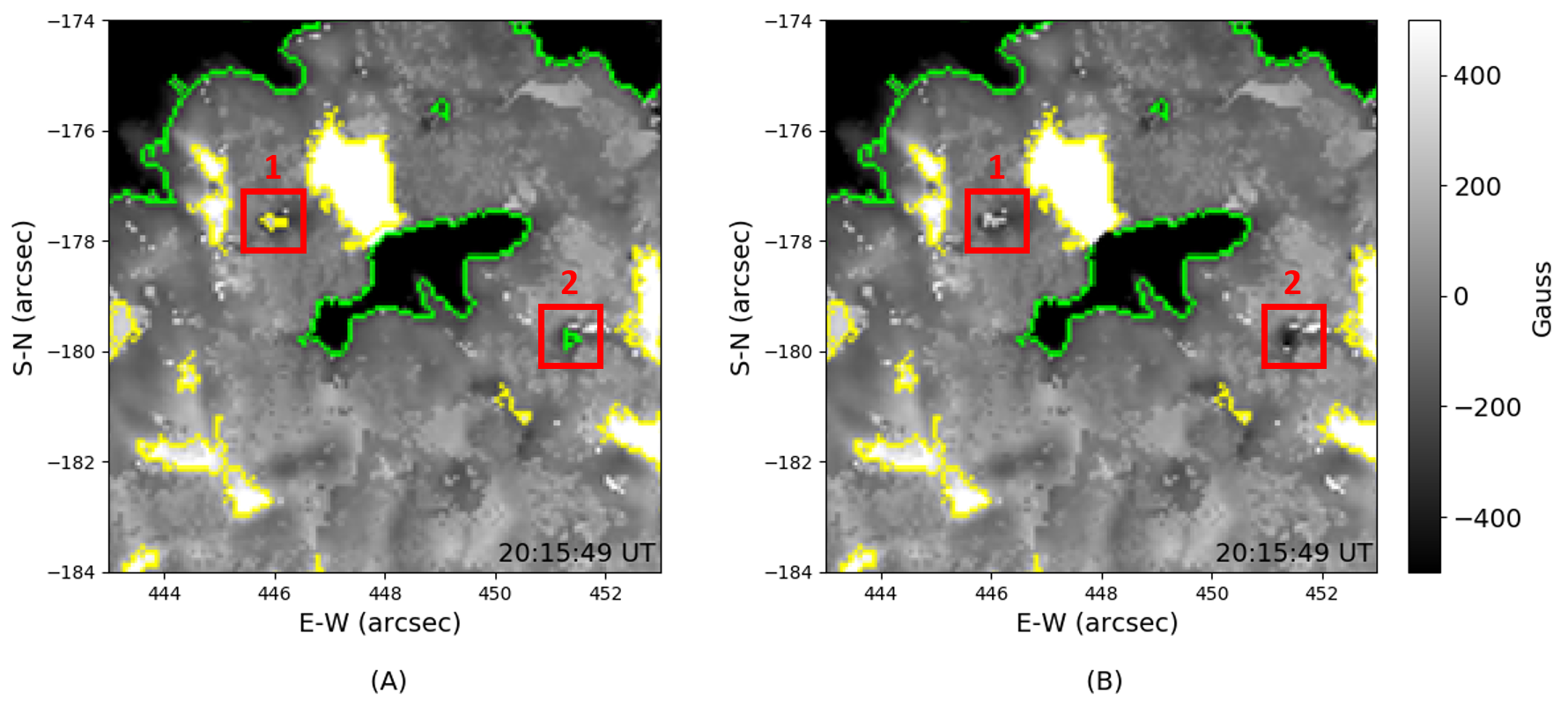}
	\caption{Illustration of the magnetic flux elements detected by SolarUnet but not found by SWAMIS 
		on the testing magnetogram from AR 12665 collected on 2017 July 13 20:15:49 UT.
		(A) SolarUnet identifies a positive feature highlighted by yellow bordering color and a negative feature 
		highlighted by green bordering color where the two highlighted features are enclosed by red square boxes numbered by 1 and 2 respectively. 
		(B) SWAMIS does not find the two features as no bordering color is shown inside the red square boxes numbered by 1 and 2 respectively.
		Time in UT is at the bottom right of each image.}
	\label{fig:comparison}
\end{figure}

It should be pointed out that, although the U-shaped network (i.e., the deep learning model) 
in SolarUnet gains knowledge from the training data prepared by SWAMIS, 
the model is able to generalize learned features to more generic forms.
In our work, the model gains knowledge from the training images in quiescent solar regions collected on 2018 June 07 and
uses the acquired knowledge to make predictions on unseen testing images from an active region (NOAA AR 12665) collected on 2017 July 13.
With the generalization and inference capability, the model 
may discover new magnetic flux elements not found by the SWAMIS method.
For example, with the filter size threshold of SolarUnet fixed at 10 pixels, 
SolarUnet detected two opposite-sign features not found by SWAMIS
on the testing image (magnetogram) from
AR 12665 collected on 2017 July 13 20:15:49 UT.
Figure \ref{fig:comparison}(A) highlights these two features;
Figure \ref{fig:comparison}(B) shows that the two features were not found by SWAMIS.

\begin{figure}
	\epsscale{1}
	\plotone{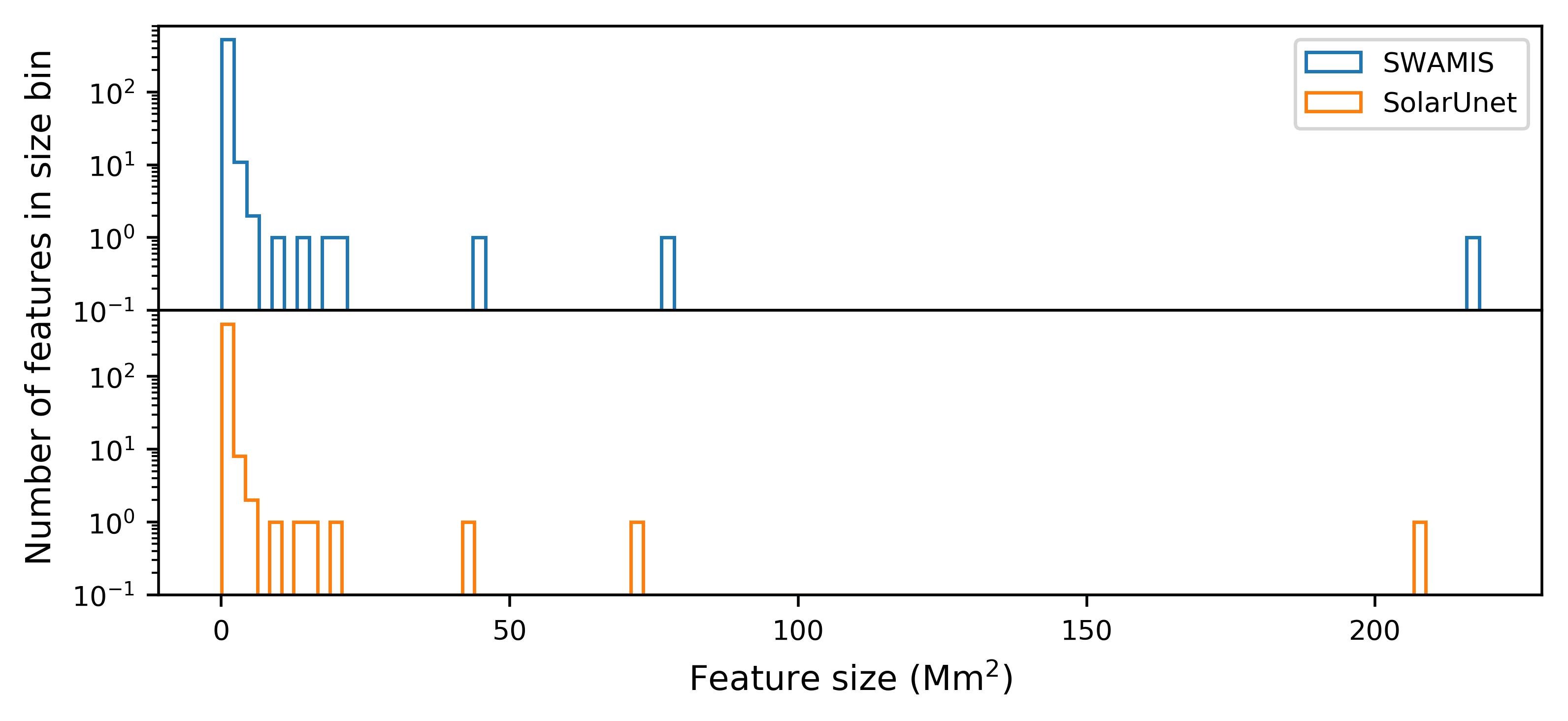}
	\caption{Magnetic feature size distributions as derived by SWAMIS (represented by blue color) and SolarUnet (represented by orange color) 
                on the testing magnetogram from AR 12665 collected on 2017 July 13 20:15:49 UT. 
                SolarUnet agrees mostly with SWAMIS on the feature size distributions.}
	\label{fig:size}
\end{figure}

Figure \ref{fig:size} compares the feature size distributions of SWAMIS and SolarUnet
on the testing image (magnetogram) where the features had at least 2 pixels (0.007242 Mm$^{2}$).\footnote{In
this and subsequent experiments, features with 1 pixel were considered as noise and excluded.}
The feature sizes of SWAMIS are represented by blue color and those of SolarUnet are represented by orange color.
It can be seen from Figure \ref{fig:size} that SolarUnet agrees mostly with SWAMIS on the feature size distributions.
To quantify this finding, we conducted the Epps-Singleton two-sample test \citep{tandf_gscs2026_177, sage_stja9_454, Gibbons2011}.
According to the test, the results of SolarUnet and SWAMIS have a significant difference when $p \leq 0.05$. 
In our case $p = 0.858 > 0.05$, and hence we conclude that the results of the two tools are similar.
Table \ref{statistics_table} shows the minimum, maximum, median, mean and standard deviation (SD)
of the feature sizes found by SWAMIS and SolarUnet, respectively.
SWAMIS detected 548 features while SolarUnet identified 543 features.
The largest magnetic feature,
which was a negative feature,
found by SWAMIS had 60213 pixels
(218.03 Mm$^{2}$).
This feature was also detected by SolarUnet, with a smaller size of 57662 pixels
(208.80  Mm$^{2}$).
This size difference occurs due to the different feature identification and tracking algorithms used by the two tools.

\begin{table}
	\caption{Summary Statistics of Feature Size and Flux Distributions as Derived by SWAMIS and SolarUnet}
	 \setlength\tabcolsep{18.5pt}
	\centering 
	\begin{tabular}{c c c c c c c} 
		\hline\hline   
		& Method & Minimum & Maximum & Median & Mean &  SD \\ [1ex]  
		\hline  
	Feature Size  & SWAMIS &0.007242 &218.03 & 0.029            & 0.94   & 10.14 \\
	(Mm$^{2}$)& SolarUnet &0.007242 &208.80 & 0.022            & 0.89   &  9.72 \\
	\hline
	Feature Flux & SWAMIS &0.009507 &1805.13       &  0.048          & 6.70  & 81.59 \\
	($10^{18}$Mx)& SolarUnet &0.011051 &1780.96 &  0.048          & 6.61  & 80.89 \\
		\hline
	\end{tabular} 
\begin{threeparttable}
	\flushleft ~\textbf{Notes.}
	\begin{tablenotes}
		\small
       \item[a] The data presented in this table are based on the testing magnetogram from AR 12665 collected on 2017 July 13 20:15:49 UT.\\
       \item[b] SWAMIS detected 548 features in the testing magnetogram. \\
       \item[c] SolarUnet identified 543 features in the testing magnetogram.
	\end{tablenotes}
\end{threeparttable}
\label{statistics_table}
\end{table}

\begin{figure}
	\epsscale{1}
	\plotone{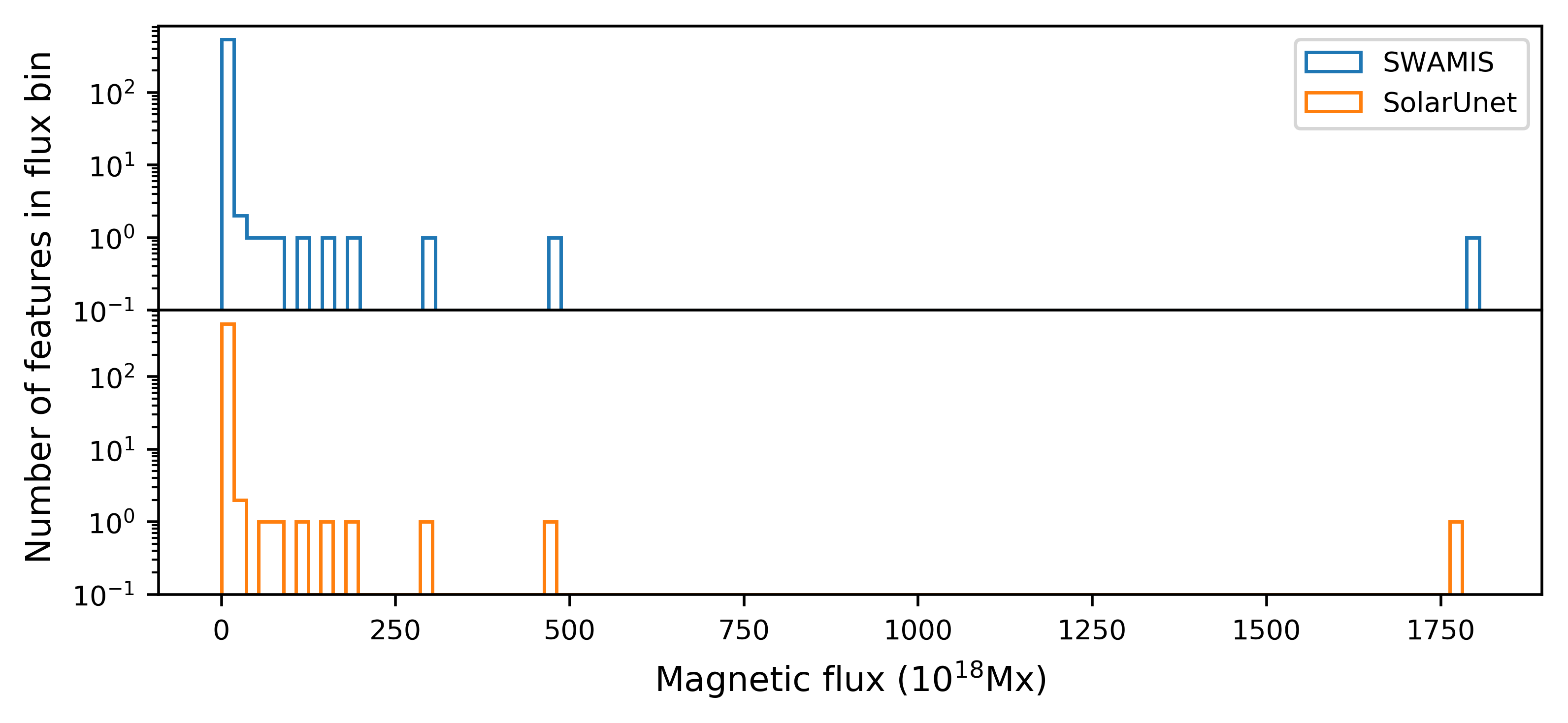}
	\caption{Magnetic feature flux distributions as derived by SWAMIS (represented by blue color) and SolarUnet
                 (represented by orange color)
                on the testing magnetogram from AR 12665 collected on 2017 July 13 20:15:49 UT.
                SolarUnet agrees mostly with SWAMIS on the feature flux distributions.}
	\label{fig:flux}
\end{figure}

Next, for each feature detected by the tools, we calculated its flux using the formula in Eq. (\ref{flux-formula}). 
Figure \ref{fig:flux} compares the feature flux distributions of SWAMIS and SolarUnet.
The results in Figure \ref{fig:flux} are consistent with those in Figure \ref{fig:size}; 
SolarUnet agrees mostly with SWAMIS on the feature flux distributions.
According to the Epps-Singleton two-sample test,
the feature flux distributions of SolarUnet and SWAMIS have a significant difference when $p \leq 0.05$. 
In our case $p = 0.983 > 0.05$, and consequently we conclude that the feature flux distributions of SolarUnet and SWAMIS are similar.
Table \ref{statistics_table} shows the minimum, maximum, median, mean and standard deviation (SD)
of the feature fluxes found by SWAMIS and SolarUnet, respectively.
The feature fluxes detected by SWAMIS ranged from 0.009507$\times$$10^{18}$ Mx to 1805.13$\times$$10^{18}$ Mx.
The feature fluxes detected by SolarUnet ranged from 0.011051$\times$$10^{18}$ Mx to 1780.96$\times$$10^{18}$ Mx.
Some of the small fluxes could be noise whiles others might be involved in
small-scale magnetic flux emergence \citep{2018ApJ...859L..26M} or small-scale magnetic flux cancellation \citep{Chen_2015}.
Similar results on feature size and flux distributions were obtained from the other magnetograms in the testing set.

To further understand the behavior of SolarUnet and compare it with SWAMIS, we performed additional
experiments to examine the lifetimes of the features identified and tracked by the two tools.
We applied SolarUnet and SWAMIS
to all of the 147 testing magnetograms mentioned in Table \ref{table_samples}.
The lifetime of a feature $X$ is defined as $X$'s disappearance time minus $X$'s appearance time.
More precisely, assuming $X$ appears in the $m$th frame and disappears after the $n$th frame 
(i.e., $X$ is not shown in the $(n+1)$th frame),
the lifetime of $X$ is defined to be $n-m+1$ frames.
Feature lifetime is strongly dependent on the feature identification and tracking algorithms employed by a tool \citep{DeForest_2007},
and can be used to measure flux turnover rate \citep{Hagenaar_2003}.

\begin{figure}
	\epsscale{1}
	\plotone{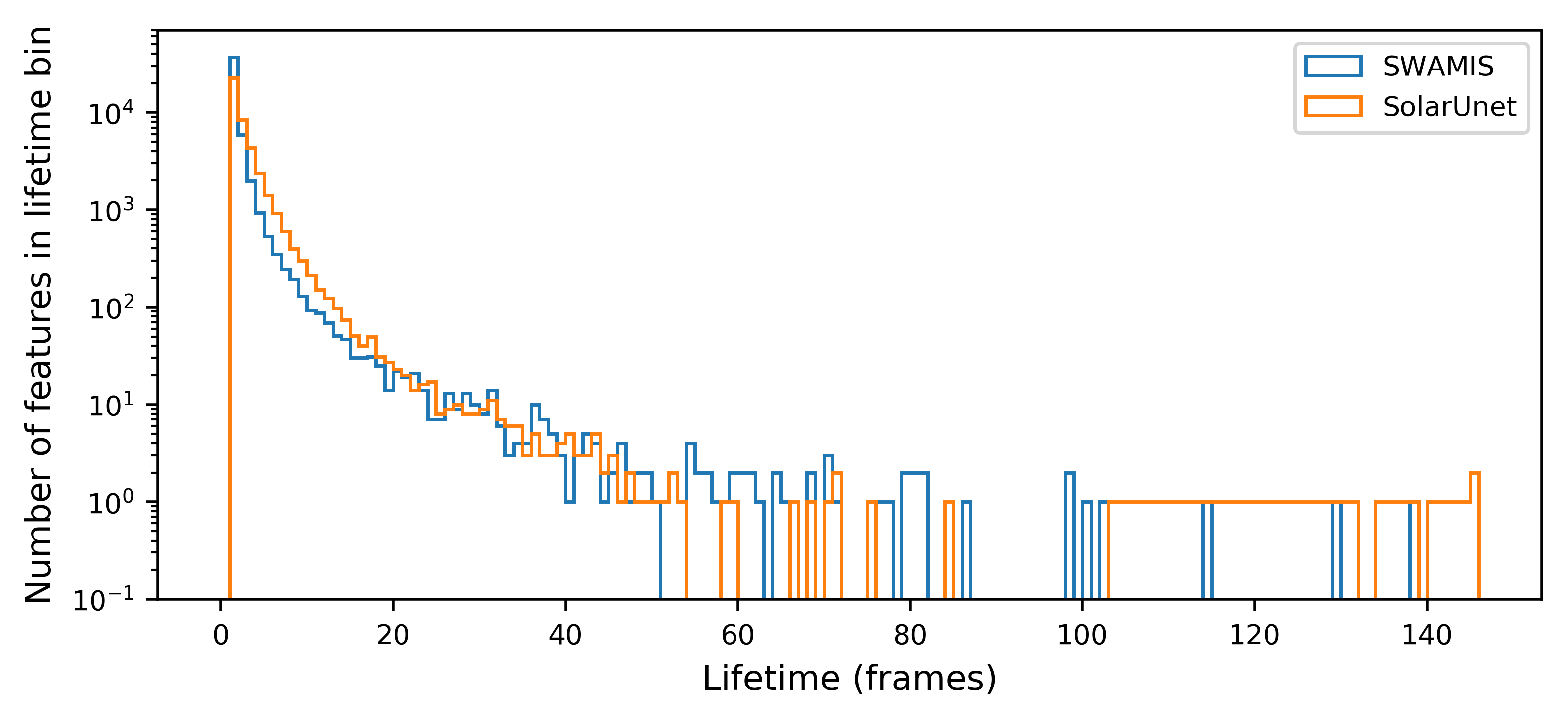}
	\caption{Feature lifetime histograms derived from SWAMIS and SolarUnet
            based on the 147 testing magnetograms (frames) from AR 12665 collected on 2017 July 13.
          SWAMIS tracks 48145 features, among which 37110 features have a lifetime of 1 frame.
	SolarUnet tracks 42470 features, among which 22657 features have a lifetime of 1 frame.
          On the other hand,  SolarUnet tracks 19813 features whose lifetimes last more than 1 frame
	while SWAMIS only tracks 11035 such features.
	SolarUnet complements SWAMIS in tracking long-lifetime features.}
	\label{fig:lifetime}
\end{figure}

Figure \ref{fig:lifetime} compares the lifetimes of features 
found by SWAMIS and SolarUnet.
SWAMIS tracked 48145 features across the 147 testing magnetograms while
SolarUnet tracked 42470 features.
The lifetimes of features found by SWAMIS ranged from 1 frame (56 seconds) to 138 frames (128.8 minutes).
The lifetimes of features detected by SolarUnet ranged from 1 frame to 147 frames (137.2 minutes).
SWAMIS tracked more short-lifetime features than SolarUnet
while SolarUnet tracked more long-lifetime features than SWAMIS.
Specifically, among the 48145 features tracked by SWAMIS,
37110 features had a lifetime of 1 frame while
SolarUnet only identified and tracked 22657 such features.
On the other hand, SolarUnet tracked 19813 features whose lifetimes lasted more than 1 frame
while SWAMIS only identified and tracked 11035 such features.
SolarUnet complements SWAMIS in tracking long-lifetime features.
We note that the training data of SolarUnet are from SWAMIS.
For those features with short lifetime in the training images,
our deep learning model may not acquire enough knowledge about them, 
and hence may miss similar features in the testing images.
This may explain why SolarUnet detects fewer short-lifetime features than SWAMIS.

\section{Discussion and Conclusions} \label{sec:conclusion}

We develop a deep learning method, SolarUnet, for tracking signed magnetic flux elements (features) 
and detecting magnetic events in observed vector magnetograms. 
We apply the SolarUnet tool to data from the 1.6 meter Goode Solar Telescope (GST) 
at the Big Bear Solar Observatory (BBSO). 
The tool is able to identify the magnetic features and 
detect three types of events, namely disappearance, merging and cancellation, in the death category and
three types of events, namely appearance, splitting and emergence, in the birth category.
We use the BBSO/GST images to illustrate how our tool works on feature identification and event detection,
and compares with the widely used SWAMIS tool \citep{DeForest_2007}.

Our main results are summarized as follows:
\begin{quote}
        1. For the testing data considered, SolarUnet agrees mostly with SWAMIS on feature size (area) and flux distributions, 
          and complements SWAMIS in tracking long-lifetime features. 
It is worth noting that because SolarUnet performs magnetic tracking through making predictions, it is
 faster than the current version of SWAMIS.
In general, SolarUnet runs in seconds on a testing magnetogram while
the current version of SWAMIS runs in minutes on the same testing magnetogram.

	2. SolarUnet is a physics-guided tool in the sense that it incorporates physics knowledge into its model and algorithms in several ways.  
First, the training data of SolarUnet are from the physics-based SWAMIS tool.
Second, in designing the loss function for the deep learning model used by SolarUnet, based on the observation that 
non-significant flux regions roughly have the same number of pixels as significant flux regions in the training set,
we adopt a binary cross-entropy loss function as defined in Eq. (\ref{magnetic loss function}) instead of the weighted cross-entropy loss function
used by the related U-Net model \citep{U-Net-2019}.
Third, in converting the binary (two-class) mask predicted by our deep learning model for a testing magnetogram
to a three-class mask with polarity information, we use the information of radial components in the vertical magnetic field image
of the testing magnetogram to reconstruct positive and negative magnetic flux regions in the predicted mask.
Lastly, by exploiting physics knowledge and based on the observational data and instruments used,
we introduce the moving distance as defined in Eq. (\ref{moving-distance})
and region of interest (ROI) as defined in Eq. (\ref{ROI}) of a magnetic flux element
to find the association of features across frames 
so as to track these features.

         3. Although SolarUnet gets training data from SWAMIS, 
         our tool may discover new features not found by the SWAMIS method.
	For example, refer to Figure \ref{fig:comparison}.
       SolarUnet may detect smaller opposite-polarity features, as shown and highlighted in Figure \ref{fig:comparison}(A), 
      near larger magnetic flux elements.  
Small-scale energy release phenomena, ranging from coronal jets down to
spicules, may be responsible for providing the upward flux of energy and
momentum for the observed heatings and flows in the corona, and may
plausibly drive the small transients in the solar wind recently
discovered by the Parker Solar Probe \citep{Parker2020}. There is
mounting evidence that these events are generated via small-scale
magnetic reconnection \citep[e.g.,][]{Samanta2019}, the photospheric
signature of which is flux cancellation involving opposite magnetic
polarities \citep{Zwaan87}. The ability of SolarUnet in detecting smaller
opposite-polarity features near larger magnetic flux elements in a 
faster manner can result in an improved determination of
magnetic reconnection rate, thus contributing to the understanding of
the mechanisms of solar coronal heating and the acceleration of the solar wind.
 
     4. The deep learning model in SolarUnet 
performs binary (two-class) classification, i.e., predicting a two-class mask, rather than three-class classification, 
i.e., predicting a three-class mask, during image segmentation. 
SolarUnet produces a three-class mask through post-processing of the predicted two-class mask
as described in item 2 above and in Section \ref{sec:overview}. 
As indicated in the machine learning literature, multiclass classification including three-class classification
often adds more noise to the loss function 
\citep[see, e.g., the Abstract of][]{Gupta-2014}, 
and it is easier to devise algorithms for binary classification
\citep[see, e.g., the Introduction in][]{Allwein-2001}.
We conducted additional experiments to compare SolarUnet with a three-class classification method.
This method trained its deep learning model using the three-class masks obtained directly from SWAMIS, and
predicted three-class masks.
Its model was the same as SolarUnet's model except that
(i) its loss function was changed from the binary cross-entropy function defined in Eq. (\ref{magnetic loss function})
to a categorical cross-entropy loss with three class labels (1, 0, $-1$);
(ii) its softmax activation function was modified to output three-class masks.
The three-class classification method used the same tracking algorithms as described in Section \ref{sec:magnetic tracking} for magnetic tracking and event detection.
The results of the three-class classification method were not as good as those of SolarUnet.
For example, the feature size distribution obtained from the three-class classification method was 
significantly different from the feature size distribution obtained from SWAMIS with $p =  0.025 \leq 0.05$ 
according to the Epps-Singleton two-sample test on the testing magnetogram from AR 12665 collected on 2017 July 13 20:15:49 UT.		
			
\end{quote}

Based on our experimental results, we conclude that the proposed SolarUnet
should be considered a novel and alternative
method for identifying and tracking magnetic flux elements. 
More testing of the method, using different training and test data, should be performed.
With the advent of big and complex observational data gathered from diverse instruments
such as BBSO/GST and the upcoming Daniel K. Inouye Solar Telescope (DKIST),
it is expected that the physics-guided deep learning-based
SolarUnet tool will be a useful utility for processing and analyzing the data.

We thank the referee for very helpful and thoughtful comments.
We also thank the BBSO/GST team for providing the data used in this study. 
The BBSO operation is supported by the New Jersey Institute of Technology
and U.S. NSF grant AGS-1821294. 
The GST operation is partly supported by the Korea Astronomy and Space Science Institute, 
the Seoul National University, and the Key Laboratory of Solar Activities of the Chinese Academy of Sciences (CAS) 
and the Operation, Maintenance and Upgrading Fund of CAS for Astronomical Telescopes and Facility Instruments. 
Portions of the analysis presented here made use of the Perl Data Language (PDL),
which has been developed by K. Glazebrook, J. Brinchmann, J. Cerney, 
C. DeForest, D. Hunt, T. Jenness, T. Lukka, R. Schwebel, and C. Soeller 
and can be obtained from \url{http://pdl.perl.org}.
This work was supported by U.S. NSF grants AGS-1927578 and AGS-1954737.
C.L. and H.W. acknowledge the support of NASA under grants NNX16AF72G, 
80NSSC18K0673, and 80NSSC18K1705.

\facilities{Big Bear Solar Observatory}

\appendix

\noindent
Here we explain the technical terms used in describing our deep learning model (i.e., the U-shaped convolutional neural network). \\

\noindent
{\em Encoder} is a neural network, which takes an input image and generates a high-dimensional vector
that is an abstract representation of the image \citep[see Chapter 8.5.2 in][]{Aggarwal-2018}.
By using the encoder, our model can better understand the content and context of the image.\\

\noindent
{\em Decoder} is a neural network, which takes a high-dimensional vector and
generates a segmentation mask \citep[see Chapter 8.5.2 in][]{Aggarwal-2018}.
By using the decoder, our model can recover the spatial information in the input image.\\

\noindent
{\em  Bottleneck}, also known as the ``compressed code" \citep[see Chapter 8.5.2 in][]{Aggarwal-2018}, 
 is a layer with less neurons than the layer below or above it \citep{bottleneck}.
In general, it can be used to obtain a representation of the input with reduced size (dimensionality).
In our model, bottleneck mediates between the encoder and the decoder. \\

\noindent
{\em Convolution layer} contains multiple kernels where 
a kernel is a matrix whose elements (weights) need to be learned from training data
\citep[see Chapter 9 in][]{Goodfellow-et-al-2016}. 
Each kernel is multiplied with an image vector $X$ (via element-wise multiplications)
to produce a new image vector that contains 
only the important information in $X$ \citep[see Chapter 8 in][]{Aggarwal-2018}.\\

\noindent
{\em Max pooling layer} reduces the size of an image vector $X$ while 
retaining only the important information in $X$
\citep[see Chapter 8.2 in][]{Aggarwal-2018}.\\

\noindent
{\em Up-convolution layer}, containing learnable parameters (weights),
increases the size of an image vector $X$.
This layer, also called an upsampling \citep{FCN-2017} or deconvolution layer,
can recover the spatial information in $X$  \citep[see Chapter 8.5.2 in][]{Aggarwal-2018}. \\

\noindent
{\em Softmax activation function} converts a vector of $k$ real values to a vector of $k$ real values that sum to 1
\citep[see page 14  in][]{Aggarwal-2018}. 
Softmax is useful because it converts the scores in the vector to a normalized probability distribution, which can be displayed to a user.
In our model, softmax is used to output the class label (1 vs. $-1$ or non-significant flux vs. significant flux) of each pixel. \\

\noindent
{\em Rectified linear unit (ReLU)} employs an activation function $f(x)$, defined as $f(x) = \max(0, x)$,  
where $x$ is the input to a neuron, $f(x) = x$ if $x \geq 0$ and $f(x) = 0$ otherwise
\citep[see Chapter 1.2 in][]{Aggarwal-2018}.
It is easy to train a model that uses ReLUs, which often achieves good performance.

\bibliographystyle{aasjournal}
\bibliography{reference}

\end{document}